\documentclass[pdftex,twocolumn,epjc3]{svjour3}          

\RequirePackage[T1]{fontenc}

\smartqed  

\RequirePackage{graphicx}
\RequirePackage{mathptmx}      
\RequirePackage{flushend}
\RequirePackage[colorlinks,citecolor=blue,urlcolor=blue,linkcolor=blue]{hyperref}

\begin{document}

\title{P stabilizes dark matter and with CP can predict leptonic phases}


\author{Ravi Kuchimanchi\thanksref{e1,addr1}}

\thankstext{e1}{e-mail: raviaravinda@gmail.com}

\institute{19 Idlewild Street, Bel Air, Maryland 21014, USA\label{addr1}}

\date{ }

\maketitle

\begin{abstract}
We find that spontaneously broken parity ($P$) or left-right symmetry
stabilizes dark matter in a beautiful way.   If dark matter has a non-real intrinsic parity $\pm i$ (e.g. Majorana fermions), parity can ensure that it cannot decay to all normal particles with real intrinsic parities.  However if Majorana couplings are absent either in the Lepton or the dark sector,  $P$ symmetry can be redefined to remove relative non-real intrinsic phases. It is therefore predicted that neutrinos and dark matter fermions must have Majorana masses if dark matter is stable due to parity. We also consider vectorlike  doublet fermions with intrinsic parity $\pm i$.  Strong $CP$ problem is solved  by additionally imposing $CP$. Leptonic $CP$ phases  vanish at the tree level in the minimal strong $CP$ solving model, which is a testable prediction. Experimentally if leptonic $CP$ phases are not found (they are found to be consistent with $0$ or $\pi$) it can be evidence for the type of models in this work where $CP$ is spontaneously or softly broken and there is also a second hidden or softly broken symmetry such as $P$, $Z_2$ or $Z_4$. However leptonic $CP$ violation can be present in closely related or some non-minimal versions of these models.
\end{abstract} 

\section{Introduction} 
Astronomical observations of  galactic rotation curves~\cite{Oort32} and velocity distribution of galaxies in clusters~\cite{Zwicky:1933gu}, smallness of anisotropies in the Cosmic Microwave Background radiation~\cite{Jarosik:2010iu}, and in a striking manner the Bullet Cluster~\cite{Clowe:2006eq}, have all provided significant evidence that there is 5 times more matter in the universe that interacts gravitationally than is visible.  The dominant thinking is that dark matter is comprised of a new non-baryonic particle, is electrically neutral and therefore almost transparent. 

If dark matter is abundantly present it must be very stable. The standard approach to prevent it from decaying to normal matter  is to introduce an unbroken $Z_2$ symmetry whereby the dark matter particles are odd under $Z_2$ while normal matter particles are even.  This then implies that the \emph{lightest} $Z_2$ odd particle is stable.  

However introducing $Z_2$ only for the purpose of stability is unsatisfactory as it does not provide greater insight (see for example~\cite{Hambye:2010zb}).  Moreover there is considerable arbitrariness in model building as there are many ways to introduce it.  Therefore there is a need for a deeper approach to the problem of stability of dark matter. 


In this work we first prove at a  fundamental level that parity ($P$) can stabilize dark matter, though it is spontaneously broken.  Illustrating this with  left-right symmetric  dark sector  we show that  $Z_2$ that is usually invoked to stabilize dark matter emerges as an automatic symmetry.  We find that if dark matter is stabilized by parity, then  dark fermions and neutrinos must have Majorana masses and there is no conserved dark charge or lepton number.  The seesaw mechanism on neutrinos maybe thus related to dark matter stability.

Since we require only $P$  to stabilize dark matter we do not depend on stability ideas that use gauge symmetries such as $U(1)_{B-L}$ or $SO(10),$ that would restrict us to multiplets with specific $B-L$ quantum numbers assigned so that R-parity or matter-parity  are automatic symmetries~\cite{PhysRevD.34.3457,Martin:1992mq,2010PhRvD81a5002K}. Our work is also distinct from models  requiring additional symmetries along with $P$ to stabilize dark matter~\cite{Guo:2008si} or that look at viability of explaining dark matter issues in the minimal left-right model itself without additional fields~\cite{Nemevsek:2012cd}. 

$P$ is well motivated not only on aesthetic grounds and because it is a discrete space-time symmetry,  but also as is well known $P$ requires that right-handed neutrinos must exist (and thus predicts that neutrinos have masses and mixing, as is now established by experiments).  Moreover  as was shown in~\cite{Kuchimanchi:2010xs}, $P$ along with  $CP$ (equivalent to another discrete space-time symmetry, time-reversal $T$ due to $CPT$ theorem) solves the strong $CP$ problem in left-right symmetric models with the addition of a vectorlike quark family.  In~\cite{Kuchimanchi:2012xb} it was noted that leptonic $CP$ violating phases can vanish in  models such as in~\cite{Kuchimanchi:2010xs}.  

In this work we show that even after inclusion of  dark matter,  leptonic $CP$ phases vanish at tree level in the minimal strong $CP$ solving model.     We also find more generally that $CP$  with an additional symmetry (such as $P$ or $Z_2$) can solve the strong $CP$ problem and predict the absence of leptonic $CP$ phases. 
If leptonic $CP$ violating phases are not detected at the sensitivity of experiments such as in~\cite{Huber:2009cw,Coloma:2012wq,Bayes:2012at} that are being planned or underway, it will hint at $CP$ being broken spontaneously (or softly) and there being a second hidden symmetry such as $P, Z_2$ or $Z_4$ as well in nature. It is interesting that a global analysis of neutrino data~\cite{2012PhRvD..86a3012F} finds possible hints for the Dirac leptonic phase to be close to the $CP$ conserving value of $\pi$, though the experimental margin of error is still very large.

The paper is organized as follows.  In section~\ref{sec:Pstabilizes} we provide a quantum mechanical argument to show that if there are relative non-real intrinsic parity phases, $P$ can stabilize dark matter though it is spontaneously broken.  In sections~\ref{sec:majorana} and~\ref{sec:xparticles} we show in specific left-right symmetric models that a relative imaginary intrinsic parity phase  cannot be rotated away by redefinitions of $P$ operator. We thus establish  in a basis independent manner that $P$ can stabilize the dark sector and predict that neutrinos and dark neutral particles must have Majorana masses.  We still need to show that in the  dark sector stabilized by $P$, the \emph{neutral} particle is stable, and we do this in section~\ref{sec:masses} by studying different regions of parameter space.  In section~\ref{sec:strongCP} we show that if the Lagranigain is invariant under both discrete space-time symmetries $P$ and $CP$,  we can simultaneously solve the strong $CP$ problem, have stable dark matter and predict the absence of leptonic $CP$ violation without requiring any other symmetry.  We also show more generally that $CP$ with an additional symmetry can predict the absence of leptonic $CP$ phases. 
 
\section{Parity stabilizes dark matter}
\label{sec:Pstabilizes}
We first argue that \emph{Parity (P)  can stabilize dark matter, even if it is spontaneously broken.}  
Let $P$ be a good symmetry of the Lagrangian (such as in the left-right symmetric model) and there be Higgs fields $\Delta_L$ and $\Delta_R$ that transform under $P$ as $\Delta_L(x,t) \leftrightarrow \Delta_R(-x,t)$. Note that indices $L$ and $R$ on scalar fields are just labels.   $P$ is spontaneously broken (or hidden)  when the neutral component $\Delta^o_R$ picks a constant vacuum expectation value (VEV) such that $\left<\Delta^o_R\right> >> \left<\Delta^o_L\right>$.   However applying $P$  twice it is easy to check that under $P^2$,  $\Delta_L(x,t) \rightarrow \Delta_L(x,t)$,  $\Delta_R(x,t) \rightarrow \Delta_R(x,t)$ and it follows  that though $P$ is broken, $P^2$ \emph{remains unbroken} by these VEVs. Note that we have used the fact that since $P$ is a good symmetry of the Lagrangian, so is $P^2$.

Classically $P^2$ (space inversion followed by space inversion) returns a system to its original state.  But Quantum Mechanically this needs to be true only up to a phase.  That is there can exist states $\psi_\alpha$ such that, $P^2 \psi_\alpha = e^{i\phi_\alpha} \psi_\alpha$ since only $|\psi_\alpha|^2$ is physically observable and not the eigenstate $\psi_\alpha$ itself. Under $P^2$ different quantum fields can pick up different phases characterized by $e^{i\phi_\alpha}$.

Note that $\eta \equiv \pm e^{i\phi_\alpha/2}$ is called intrinsic parity as it is the parity of underlying $P$ eigenstates~\cite{PhysRev.88.101}. Hence we use  \emph{intrinsic parity squared} for  the $P^2$ eigenvalue $\eta^2$.

Since $P^2$ is conserved, the lightest particle $\chi$ with intrinsic parity squared 
$\eta_\chi^2 \neq 1$ (\emph{we identify $\chi$ with dark matter}) cannot decay into a final state with intrinsic parity squared $1$.  That is into a final state consisting of particles that all have intrinsic parity squared 1 (or normal matter). This ensures the stability of dark matter.

However if $P$  is redefined to remove all complex $\eta,$  in that basis $P$ cannot explain dark matter stability, which must then be due to another symmetry.  Since the redefined $P$ operator has to be a symmetry this restricts the possible redefinitions to  $P \rightarrow PU$, where unitary transformation $U$ is a multiplicative symmetry of the Lagrangian~\cite{feinberg1959phase,DeGraaf1969142}.  As we shall see in the next section if $\eta_\chi = \pm i$, dark matter fermions  can have Majorana masses. Neutrinos with $\eta_L = \pm 1$ can also have Majorana masses. Due to presence of Majorana terms (couplings that give rise to Majorana masses), along with all the usual terms consistent with parity and gauge symmetry, there is not enough symmetry to remove the purely imaginary relative intrinsic parity phase by $P$ redefinitions. In this case, as shown in the next section in a basis indepedent manner,  $P$ stabilizes dark matter.


On the other hand, for non-real $\eta_\chi \neq \pm i$, Majorana terms are disallowed by $P$ for dark sector  and $\eta_\chi$ can be made real through a redefinition of parity symmetry.   In this  case $P$ cannot stabilize dark matter.

\section{LR Symmetry, Majorana mass and dark matter stability}
\label{sec:LRsymmetry}
We now consider the well known left-right symmetric group~\cite{PhysRevD.10.275,PhysRevD.11.566,Senjanovic:1975rk} $G_{LR} \equiv SU(3)_c \times SU(2)_L \times SU(2)_R \times U(1)_{B-L} \times P$ which is the most popular and economical way to restore parity as a good symmetry of the Lagrangian so that it can be spontaneously broken. We consider the usual  Higgs content suitable for symmetry breaking, namely, $SU(2)_R$  Higgs fields (either doublet $H_R$ and its parity partner $H_L$ as in section~\ref{sec:majorana}, or triplet $\Delta_R$ and its parity partner $\Delta_L$ as in section~\ref{sec:xparticles}), and the bidoublet $\phi$. 



The matter content  consists of the usual 3 generations of quarks and leptons represented by $Q_{iL}, Q_{iR}, L_{iL}$ and $L_{iR}$  (with $i=1,2,3$) such that under $P$, space-time coordinates $(x,t) \rightarrow (-x,t)$,  the gauge bosons of  $SU(2)_L \leftrightarrow SU(2)_R$ and, $Q_{iL} \leftrightarrow Q_{iR}, \ L_{iL} \leftrightarrow L_{iR}, \ H_L \leftrightarrow H_R$ (or $\Delta_L \leftrightarrow \Delta_R$)  and $ \phi \rightarrow \phi^\dagger$.  

The above is the usual parity transformation commonly used in left-right symmetric models.  By applying $P$ twice we note that all the above states are unchanged and therefore we have $\eta^2 = 1$ for all the usual particles. 

We recall that in the Left-Right model the representation of particles is as follows:
\begin{equation}
\begin{array}{cc}
L_{1L,1R}=\left(\begin{array}{c}
\nu_e \\ e^-
\end{array} \right)_{L,R},
& \phi = \left(\begin{array}{cc}
\phi^o_1 & \phi^+_2 \\
\phi^-_1 & \phi^o_2
\end{array}
\right), \\

H_{L,R}=\left(\begin{array}{c}
H^+_{L,R} \\ H^o_{L,R}
\end{array} \right), &
\Delta_{L,R} = \left(\begin{array}{cc}
\delta^+_{L,R} / \sqrt{2} & \delta^{++}_{L,R} \\
\delta^o_{L,R} & - \delta^+_{L,R} /\sqrt{2}
\end{array}
\right)
\end{array}
\end{equation}
where as an example of the Lepton and Quark doublets we have shown the first generation Lepton doublets $L_{1L}$ and $L_{1R}$.

$\left<H^o_R\right> >> \left<H^o_L\right>$ (or $\left<\delta^o_R\right> >> \left<\delta^o_L\right>$)  breaks $G_{LR}$ to the standard model. $P$ is also broken, but $P^2$ remains unbroken.  $\left<\phi^o_{1,2}\right>$ cause the electro-weak symmetry breaking and provides usual Dirac masses to fermions through Yukawa couplings such as 
\begin{equation}
\bar{L}_{iL}  \left(h_{ij} \phi + \tilde{h}_{ij} \tilde{\phi}\right) L_{jR} + H.c.
\label{eq:dirac}
\end{equation}
(with $\tilde{\phi} = \tau_2 \phi^\star \tau_2$). As is well known in left-right models, Yukawa matrices (with matrix elements $h_{ij}$ and $\tilde{h}_{ij}$) involving the bi-doublet are  Hermitian due to parity.   Also all terms in the Higgs potential consistent with parity,  such as $\mu_H H^\dagger_L \phi H_R,$ \ $\tilde{\mu}_H H^\dagger_L \tilde{\phi} H_R,$ \ $\mu^2 \tilde{\phi}^\dagger \phi$, $\beta_1 Tr ( \Delta^\dagger_L \phi \Delta_R\phi^\dagger)$ and $\alpha_2 [Tr(\phi^\dagger \tilde{\phi})$ $Tr (\Delta^\dagger_L \Delta_L)$ $+ \ Tr(\phi \tilde{\phi}^\dagger) Tr (\Delta^\dagger_R \Delta_R)]$   (with their Hermitian conjugates, and  $\mu_H, \ \tilde{\mu}_H, \ \mu^2, \ \beta_1$ real due to $P$), are present if the corresponding Higgs field is present. This ensures that there are no additional symmetries in the model than what we will consider. This is important to note since if they are present, symmetries can be used to redefine the $P$ operator. The Higgs potential for minimal left-right symmetric model is given in~\cite{Duka:1999uc} and we do not write all the terms here.

\subsection{Singlet Majorana fermion}
\label{sec:majorana}
We include a Majorana fermion $X_M$ which is a singlet of the gauge group, and all the fermionic and Higgs fields in the model  are as in Table~\ref{Tab:Table0}. Since Majorana fermions are their own antiparticles, their intrinsic parity $\eta_{X_M} = \pm i$~\cite{Weinberg:1995mt}. Under $P,$ $X_M \rightarrow i \gamma_o X_M $ so that under $P^2$, $X_M \rightarrow -X_M$.  Note that we have assigned $\eta = 1$ to all other particles including  $\eta_L = 1$ to leptons.

\begin{table}[t]
\centering
\begin{tabular}{|cc|c||c|c|c||c|c|c|}
\hline
Group & $Q_{iL}$  & $Q_{iR}$ & $L_{iL}$ & $L_{iR}$ & $X_M$& $H_L$ &$H_R$ &$\phi $   \nonumber \\
\hline 
$SU(3)_c $& 3 & 3 & 1 & 1&1&1&1&1  \nonumber \\
$SU(2)_{L} $ &  2 &  1 & 2 & 1 &1&2&1&2\nonumber \\
$SU(2)_R$ & 1 & 2 & 1 & 2&1&1&2&2 \nonumber \\
$B-L$ & 1/3 & 1/3 & -1 & -1&0&1&1&0 \nonumber \\
\hline 
\end{tabular}
\caption{Left-right symmetric model with addition of a Majorana fermion $X_M$. $i = 1$ to $ 3$ correspond to the usual 3 chiral families of quarks and leptons.}
\label{Tab:Table0}
\end{table}

The most general $P$ invariant mass and Yukawa term involving $X_M$  is just $m_M X_M^T C X_M$.  Yukawa coupling terms such as  $\bar{L}_{iL} \tilde{H}_L X_M$ (with $\tilde{H}_{L} \equiv i\tau_2 H^*_{L}$) can cause the decay of the Majorana fermion and are permitted by gauge invariance. However they are odd under $P^2$ and are thus absent due to parity, making the $X_M$ particle stable. 

We allow Majorana type neutrino masses and leptonic mixing to arise from the parity symmetric non-renormalizable Majorana term of form 
\begin{equation}
 \left({\frac{f_{ij}}{M}} \right)\left( L_{iL} H_L L_{jL} H_L  + L_{iR} H_R  L_{jR}H_R \right) + H.c.
 \label{eq:doublet_maj}
 \end{equation}
  with $i,j = 1, 2, 3$, where for the ease of notation we have suppressed $C,$ Pauli matrix $\tau_2$ and transposes.  Due to this term there is no  lepton number symmetry ($U(1)_L$ acting only on leptons) in the model using which the parity symmetry can be redefined to remove the relative imaginary intrinsic parity phase between the leptons and  $X_M$. 



However the Lagrangian has the multiplicative symmetry ~\cite{feinberg1959phase,DeGraaf1969142} $U(1)_{B-L}$, which can be used to redefine P.   Under $P\rightarrow PU(1)_{B-L}$, the intrinsic parities transform as $\eta \rightarrow e^{i(B-L) \theta} \eta$, where  $\theta$ is any angle.  Note from Table~\ref{Tab:Table0} that  $\tilde{H}_{L,R}$ and $L_{iL,iR}$ have the same $B-L$ charge, and their intrinsic parities will remain equal.  The bi-doublet $\phi$ and $X_M$ are $B-L$ singlets.  Their intrinsic parity will be unchanged and remain real and purely imaginary respectively.   Thus there will necessarily be a relative non-real intrinsic parity phase either between $\phi$ and $H_{L,R}$ or between the leptons and $X_M$, that cannot be removed by $P$ redefinition. Moreover using the redefined intrinsic parities  it is easy to see that $\bar{L}_{iL} \tilde{H}_L X_M$ remains odd under  $P^2$, and therefore the stability of $X_M$ is established as being due to parity in a basis independent manner.

The only other multiplicative symmetry present is $Z_2$, since  $P^2$  implies that the Lagrangian is invariant under $X_M \rightarrow - X_M$.  However redefining the parity operator to be $PZ_2$ does not help in removing complex phases or change the above conclusion.

Note that if we originally provided intrinsic parity $\eta_L =  i$ instead of $\eta_L =  1$ to the Leptons (so that $\eta_L^2 = \eta_{X_M}^2$), it would have led to a different Lagrangian. That the two choices of intrinsic parity lead to different physics and therefore correspond to inequivalent Lagrangians is shown in~\ref{appendix}. 


In this example, since the dark matter $X_M$ particles do not have any interactions with ordinary matter (other than gravitationally), they cannot be produced, but must be present in the universe as an initial condition.  Dark matter abundance would then have to be understood anthropically -- much more than what is observed  will over close the universe, and too little may not help with galaxy formation.

It is possible to introduce non-renormalizable terms that also involve $X_M$, so that they  couple to other particles. Such terms would once again have an even number of $X_M$ due to $P^2$ and therefore $X_M$ remains stable.  

We now discuss the stability issue in the renormalizable left-right symmetric model with vectorlike doublet fermions.   

\subsection{Vectorlike doublet fermions with intrinsic parity \it{i} \normalfont}  
\label{sec:xparticles}
We  consider the more interesting minimal Left-Right model with triplet Higgses $\Delta_{L,R}$ (see Table~\ref{Tab:Table}) that enable neutrino masses to be generated via renormalizable terms. Under $P$ we let $\Delta_L \leftrightarrow \Delta_R$ as is usual in Left-Right symmetric models. Majorana masses are generated for the neutrinos  due to the $P$ invariant term 
\begin{equation} if_{ij} (L^T_{iL} C \tau_2 \Delta_L L_{jL} +  L^T_{iR} C \tau_2 \Delta_R L_{jR}) + H.c.
\label{eq:maj}
\end{equation} 
where as before the  leptons  have been assigned $\eta_L= 1$, which also determines the plus sign in brackets equation~(\ref{eq:maj}).   We include a fermionic particle $X_L$ which is a doublet of  $SU(2)_L$.  Due to parity, $X_R$ is automatically present. $X'_{R,L}$ are added to cancel chiral anomalies. $X_{L,R}$ consists of a singly charged and neutral fermion just like the Lepton doublets and can be represented as $X^T_{L,R} = (X^o \ X^-)_{L,R}$ and likewise for $X'_{L,R}$. 

We assign the `$X$-particles',  $X_{L}, X_{R}, X'_L$ and $X'_R$,  with non-real intrinsic parity so that under $P$,
$X_{L,R} \rightarrow \eta_X X_{R,L}$ and $X'_{L,R} \rightarrow \eta_X X'_{R,L}$ with $|\eta_X| = 1$ and $\eta_X^2 \neq 1$.  
The most general parity symmetric Yukawa and mass terms involving the $X$-particles 
are
\begin{eqnarray}
 L =   \bar{X}_L \left( h \phi + \tilde{h} \tilde{\phi} \right) X_R +  \bar{X}'_L \left( h' \phi^\dagger + \tilde{h}' \tilde{\phi}^\dagger \right) X'_R  + \nonumber \\ if ( X^{T}_L C \tau_2 \Delta_L X_L  - X^{T}_R C \tau_2 \Delta_R X_R) + ~ \nonumber \\  if' ( X'^{T}_L C \tau_2 \Delta_R X'_L - X'^{T}_R C \tau_2 \Delta_L X'_R)  + \nonumber \\ M_X \bar{X}_L X'_R + M^\star_X  \bar{X}'_L X_R + H.c. ~
\label{eq:dark}
\end{eqnarray}
\begin{table}[t]
\centering
\begin{tabular}{|cc|c||c|c||c|c|c|}
\hline
Group & $Q_{iL}$  & $Q_{iR}$ & $L_{iL}$ & $L_{iR}$ & $\Delta_L$ &$\Delta_R$ &$\phi $   \nonumber \\
 & & & $X_L,X'_R$ & $X_R,X'_L$ &&&  \nonumber \\
\hline 
$SU(3)_c $& 3 & 3 & 1 & 1&1&1&1  \nonumber \\
$SU(2)_{L} $ &  2 &  1 & 2 & 1&3&1&2\nonumber \\
$SU(2)_R$ & 1 & 2 & 1 & 2&1&3&2 \nonumber \\
$B-L$ & 1/3 & 1/3 & -1 & -1&2&2&0 \nonumber \\
\hline 
\end{tabular}
\caption{Renormalizable minimal LR symmetric model with  $X$ particles. $i = 1$ to $ 3$ correspond to the usual 3 chiral families of quarks and leptons.}
\label{Tab:Table}
\end{table}
\vspace{0.1cm}
where as before under $P, \phi \rightarrow \phi^\dagger$ (i.e. $\eta^2_\phi = 1$).  $P$ implies that $h, h', \tilde{h}, \tilde{h}'$ are all real for any choice of $\eta_X,$ real or complex. This can be seen by comparing the relevant terms with the Hermitian conjugates and using $P$. 

Terms that could couple $X$ particles to the usual leptons are automatically absent due to parity. 
For example under $P^2$ terms containing an odd number of $X$-particles  such as $\bar{L}_{iL} \phi X_R  \rightarrow \eta_X^2\bar{L}_{iL} \phi X_R $ and are not invariant for $\eta_X^2 \neq 1$. If $\eta_X$ was $\pm1$ such a term would be present in the $P$ symmetric form $\bar{L}_{iL} \phi X_R \pm \bar{L}_{iR} \phi^\dagger X_L$. 

Using the $P$ transformations it is easy to check that Majorana terms with non-zero $f,f'$ are allowed in equation~(\ref{eq:dark}) if $\eta_X =\pm i$. The minus sign in the brackets for these terms  is determined by $P$ invariance with  $\eta_X^2=-1$, and is opposite of the plus sign in brackets of $f_{ij}$ term for Leptons in equation~(\ref{eq:maj}) that was determined by $\eta^2_L=1$.  

Since Majorana terms are present both for Leptons and $X$-particles, the  maximal multiplicative symmetry of the Lagrangian that involves either of these particles is $U(1)_{B-L} \times Z_2$. Using the $B-L$ charges in Table~\ref{Tab:Table}, and following the same method as in section~\ref{sec:majorana}, we can now show that the relation  $\eta_X/\eta_L = i$  is invariant under parity redefinition $P \rightarrow PU(1)_{B-L}$.   
$Z_2$ can at the most flip the sign of this ratio.  As before, using the redefined intrinsic parities it is easy to check in a basis invariant way that terms with an odd number of X-particles such as $\bar{L}_{iL} \phi X_R$ remain odd under $P^2$.

We now consider the case where  $\eta_X \neq \pm 1, \pm i$.  In this case,  invariance under $P^2$ implies that   $f =f' = 0$ in equation~(\ref{eq:dark}) and the Majorana terms are absent for $X$-particles. This can be seen since under $P^2, X^{T}_L C \tau_2 \Delta_L X_L  \rightarrow \eta_X^4 X^{T}_L C \tau_2 \Delta_L X_L$ and cannot be present for $\eta_X^4 \neq 1$.    The Lagrangian now has a $U(1)_D$ symmetry under which $X_{L,R} \rightarrow e^{i\beta} X_{L,R}$ and $X'_{L,R} \rightarrow e^{i\beta} X'_{L,R}$ for any $\beta$.  Redefining parity to be $PU(1)_D$ implies $\eta_X \rightarrow e^{i\beta} \eta_X$.  With an appropriate choice for $\beta$ we can make $\eta_X$ real, and non-real intrinsic parities are rotated away.    In a basis with the redefined parity we would need to impose $U(1)_D$ to explain dark matter stability, and therefore dark matter is not stable due to $P$ in this case. 

Likewise if Majorana terms involving the Leptons are absent, there will be a $U(1)_L$ Lepton number symmetry using which we can rotate the intrinsic parities of the Leptons so that they become the same as those of the $X$-particles.  

Thus we predict that if dark matter is stable due to parity, there must be a relative purely imaginary intrinsic parity phase, and Majorana masses must exist both for neutrinos and for neutral $X$-particles (that we identify with dark matter particle $\chi$). This implies that the seesaw mechanism must be active, and that the dark charge and lepton number are not conserved. Experiments are currently underway to probe the nature of the neutrino and to try and establish whether they have Majorana masses. (See for example Ref.~\cite{Hall:2013,giuliani2012neutrinoless}). 

In~\ref{sec:appendixb} we show that some Dirac terms must also be present in this model.   

Since the $X$-particles are vector-like, unlike neutrinos, all neutral $X$-particles can have a large mass. The mass scale $M_X$ in equation~(\ref{eq:dark}) is independent of  the parity breaking scale set by $v_R = \left<\delta^\circ_R\right>$ and both these scales can be anywhere between the weak scale (or $TeV$) and the Planck scale. In the next section we use equation~(\ref{eq:dark}) to find the mass splitting between the charged and neutral $X-$particles and discuss the stability of the \emph{neutral} dark sector particle.

\section{Splitting of dark sector masses}
\label{sec:masses}
As we will now see using equation~(\ref{eq:dark}), in the region of parameter space $M_X >> f v_R \sim f' v_R > v_{wk}$ (where $v_{wk}$ is the weak scale), the charge-neutral dark particle  automatically has the smallest mass. In this region, ignoring the weak scale terms, the mass matrix of the charge-neutral component of the $SU(2)_R$ doublet $X$-particles ($X^o_R, X'^o_L$ and their charge conjugates) is block diagonal with $2 \times 2$ blocks of the form $\left(\begin{tabular}{cc} $-f v_R$ & $M_X$ \\ $M_X$ & $-f'^\star v_R$ \end{tabular}\right).$ Treating $fv_R$ as a perturbation, the mass eigenvalues of the charge neutral particles are split and are $M_X \pm (f + f') v_R / 2,$ if we assume that all the couplings are real.  In case they are not all real,  there may in general be a phase $\alpha$ between the terms that cannot be removed and  the splitting would then depend also on $cos \alpha$.   In any case, except for a small region near $\alpha = \pi/2$, the charge neutral $X$-particles are split such that the smaller of the eigenvalues is lower than $|M_X|$ by $O(|f v_R cos \alpha|)$.   On the other hand the $X$-particles with electric charge ($X^-_{L,R}$ and $X'^-_{L,R}$) do not receive mass corrections from $v_R$ and are only split from $M_X$ due to weak scale corrections $h v_{wk}$. Thus in the dark sector the charge neutral $X$-fermion is in fact the one with the lowest mass in a large  region of parameter space with $M_X >> f v_R \sim f'v_R > v_{wk}$. 

On the other hand if $f v_R \sim f' v_R >> M_X \geq v_{wk}$, with $f v_R >> TeV$ scale,  the charge-neutral $SU(2)_R$ doublet $X$-particles with masses $\sim O(f v_R)$ are much heavier than their $SU(2)_L$ doublet counterparts  which will have mass only slightly split from $M_X$.  However the lower mass in this case will be of the charged fermions rather than the neutral ones.  This is because the splitting for neutral components of $SU(2)_L$ doublet fermions ($X^o_L, X'^o_R$ and their charge conjugates) will have a see-saw like suppression and their masses will be $|M_X| \pm O(h^2v^2_{wk}/fv_R)$.  While the masses of the charged fermions (linear combinations of $X^-_L$ and $X'^-_R$) are $|M_X| \pm [(h + h') \kappa' + (\tilde{h} +\tilde{h}') \kappa]/2 $ as they are obtained from the mass-matrix \\ $\left(\begin{tabular}{cc}$h \kappa' + \tilde{h} \kappa$ & $M_X$ \\ $M^\star_X$ & $h' \kappa' + \tilde{h}' \kappa$ \end{tabular}\right)$. To keep the calculation simple, we have assumed that  the VEVs of $\phi$ that is, $\kappa$ and $\kappa'$ are real.  As we shall see in the next section and as shown in~\cite{Kuchimanchi:2010xs,Kuchimanchi:2012xb} if strong CP problem is solved by $P$ and $CP$ symmetries without introducing an axion, these VEVs are indeed real.  

Since the lightest particle in the dark sector is now charged it cannot serve as dark matter.  However even in this case, the charge neutral fermions can form dark matter, if their decay to charged dark particles, for example via $X^o \rightarrow X^- + e^+ + \nu_e$ is kinematically prohibited. This can  happen if the mass difference between $X^-$ and $X^o$ is less than the electron's mass.  In other words the Yukawa couplings $h, h', \tilde{h}, \tilde{h'}$ of the $X$-particles must in this case be less than or at best of the order of the Yukawa term of the electron $m_e / v_{wk}$.   Note that from now on we are  using $X^-$ and $X^o$ to generically refer to linear combinations of $X$-particles with charge $-1$ or $0$ respectively, that have the lowest mass for each charge.


Smallness of fermionic couplings does not cause a fine-tuning problem since setting  them to zero restores a symmetry of the Lagrangian~\cite{t1980naturalness}. For example if the Dirac type Yukawas terms $h = h'=  \tilde{h} = \tilde{h'} = 0, $ Eq.~(\ref{eq:dark})  is invariant under $X_R \rightarrow -X_R$ and $X'_L \rightarrow -X'_L$. The smallness of these Yukawas can be naturally understood  as being due to the small breaking of this symmetry. Likewise setting $M_X$ to zero restores the symmetry $X_{L,R} \rightarrow -X_{L,R}$ and therefore $M_X$ can be naturally small.   (Please also see~\ref{sec:appendixb}).

If $M_X$ is small enough to be within reach of LHC or future colliders, there is the exciting possibility that these charged stable dark  particles maybe produced and detected.

As the universe cools below the mass-scale of the X-particles, the slightly heavier $X^o$ (together with its anti-particle) which is kinematically stable would form the dark matter since the charged dark particles can attract one another and annihilate with their anti-particles through $X^- X^+ \rightarrow \gamma \gamma$. $X^+$ and $X^-$ would have been produced in roughly equal amounts through thermal equilibrium processes.  


 If $v_R >> v_{wk},$ $SU(2)_L \times SU(2)_R \times P$ breaks to $SU(2)_L \times U(1)_Y \times Z_2$, with $Z_2 \equiv P^2$ and the low-energy physics will be captured by the standard model group with particle content and boundary conditions determined by $P$.  An interesting paper~\cite{Arina:2011cu} that may be of relevance to our model, works out dark matter relic abundance  using $SU(2)_L \times U(1)_Y \times Z_2$ with dark vector-like iso-doublet fermions and includes Higgs triplets $\Delta_L$ so that there are Majorana terms. Without changing its dark matter phenomenology, this model can be completed in the ultraviolet by our model (with $M_X << f v_R$), 
so that $Z_2$ is not arbitrarily imposed but is identified with $P^2$. The mass-splitting   between the lightest neutral and charged dark matter particles in Ref.~\cite{Arina:2011cu} is $\sim m_e/10$, making them both stable like in our model. That the charged dark particles are also stable was not explicitly noted in~\cite{Arina:2011cu}.  

We can avoid charged stable particles for the case $M_X << f v_R$  by introducing a new scale $f'v_R \sim O(TeV) << f v_R$, so that the splitting of one of the dark neutral particles is $(h' v_{wk})^2 / (f' v_R) \sim O(10) GeV.$  The  lightest charged dark particle can naturally have a mass splitting $\leq  m_b \approx 4 GeV$, where $m_b$ is the $b$ quark mass. The neutral dark matter particle, since it is split by a greater amount, will now be the lightest and only stable one. Once again the hierarchy $f' << f$ and smallness of $M_X$ can be naturally understood as being due to an approximate symmetry of Eq.~(\ref{eq:dark}) under which  $X'_{L,R} \rightarrow i X'_{L,R}$. A recent paper~\cite{Arina:2012aj} extends the work of~\cite{Arina:2011cu} by enriching its matter content with a $SU(2)_L$ singlet neutral vectorlike dark fermion (that can be obtained from our $SU(2)_R$ vectorlike doublet)  and providing it Majorana mass of $O(TeV)$ (corresponding to our $f'v_R \sim O(TeV)$) to arrive at a similar result.  
 
Thus there are existing dark matter models that capture the physics of our model when the scale of $P$ breaking is large.   Further analysis of  dark matter phenomenology in different regions of parameter space  may be interesting and can be explored in future.



However the  models we discussed  must be extended since they suffer from the strong $CP$ problem. A way to do this  is to invoke Peccei-Quinn symmetry~\cite{PhysRevLett.38.1440} resulting in  the well known axion as a dark matter candidate. However  we now have  $X$-particles that can serve as dark sector. Moreover the strong $CP$ phase ($\bar{\theta}$) vanishes due to  $P$ itself, if it is unbroken.   This provides strong motivation to resolve the strong $CP$ problem  without an axion as shown in~\cite{Kuchimanchi:2010xs}  where $P$  itself (with $CP$) ensures that $\bar{\theta}$ is not generated at the tree level even after spontaneous (or soft)  breaking of $P$ and $CP$. That dark matter stability and solution to the strong $CP$ can be achieved  by discrete space-time symmetries $P$ and $T$ (or $CP$) is interesting.     Moreover  there is an experimentally testable prediction in the minimal model where $P$ stabilizes dark matter and with $CP$ solves the strong $CP$ problem  --  \emph{the leptonic $CP$ phases also vanish at the tree level}. 

\section{Absence of strong and leptonic CP}
\label{sec:strongCP}
In order to solve the strong $CP$ problem we impose both $P$ and $CP$ on the left-right symmetric Lagrangian and introduce a complete family of vectorlike quarks (with $Q_{4L}, Q'_R$ making a vectorlike $SU(2)_L$ doublet and $Q_{4R}, Q'_L$ making a vectorlike $SU(2)_R$ doublet) as in~\cite{Kuchimanchi:2010xs}.   Along with the $X$-particles (that can be thought of as being vectorlike and lepton-like), the fermionic and Higgs content of the model is as given in Table~\ref{Tab:Table1} .  
\begin{table}[t]
\centering
\begin{tabular}{|cc|c||c|c||c|c|c|}
\hline
Group & $Q_{iL},Q'_R$  & $Q_{iR},Q'_{L}$ & $L_{kL}$ & $L_{kR}$ & $\Delta_L$ &$\Delta_R$ &$\phi $   \nonumber \\
 & & & $X_L,X'_R$ & $X_R,X'_L$ &&&  \nonumber \\
\hline 
$SU(3)_c $& 3 & 3 & 1 & 1&1&1&1  \nonumber \\
$SU(2)_{L} $ &  2 &  1 & 2 & 1&3&1&2\nonumber \\
$SU(2)_R$ & 1 & 2 & 1 & 2&1&3&2 \nonumber \\
$B-L$ & 1/3 & 1/3 & -1 & -1&2&2&0 \nonumber \\
\hline 
\end{tabular}
\caption{Minimal strong $CP$ solving LR symmetric model with $X$ particles.  $k=1$ to $ 3$ correspond to the 3 chiral families of leptons. $i=1$ to $ 4$ include the 3 chiral families and fourth normal component of the vectorlike quark family.}
\label{Tab:Table1}
\end{table}

Note that in this model since Majorana terms such as $Q^T_{4L} \tau_2 C\Delta_{L} Q_{4L}$  violate gauge invariance and are not permitted,  the $4^{th}$ generation quarks  cannot be made stable by parity.   Since no other symmetries are imposed, they cannot belong in the dark sector but must couple to other quarks.  As we shall see these couplings generate the CKM phase in the quark sector when $CP$ is softly or spontaneously broken.   However  $X$-particles, with $\eta_X = i$ and Majorana masses belong in the dark sector, do not couple to  leptons and therefore do not generate $CP$ violation in the leptonic sector. 


We now show this more concretely.  Under $P,$ $Q_{4L}, Q'_R \leftrightarrow Q_{4R}, Q'_L$. As in~\cite{Kuchimanchi:2010xs} we impose $CP$ and break it softly by dimension 3 terms, that are $P$ symmetric namely
\begin{equation}
\sum_{i =1 \ to \ 4} M_i \bar{Q}_{iL} Q'_R + M^\star_i \bar{Q}'_L Q_{iR} +H.c.
\label{eq:M_i}
\end{equation}
  
Comparing the above  with the last two terms of equation~(\ref{eq:dark}) we see that unlike $X'_{L,R}$,  $Q'_{L,R}$ couple to all the quark  families with complex $CP$ violating couplings $M_i$. The mass of the vectorlike quarks $M \sim \sqrt{\sum |M_i|^2}$ can be any scale from just above the weak scale to the Planck scale, while the complex phases in the ratios $M_i/M$ generate the CKM $CP$ phases in the light $3 \times 3$ sector when the heavier vectorlike quarks decouple~\cite{Kuchimanchi:2010xs}.  Also, since the terms in~(\ref{eq:M_i}) do not break $P$ there is no $\bar{\theta}$ generated by them at the tree-level.

However  we note that if the scalars $\phi$ and $\Delta_{L,R}$ pick up complex VEVs they can induce leptonic CP violation.  But any complex VEVs would  break both $P$ and $CP$ and give rise to $\bar{\theta}$. Hence resolution of the strong $CP$ problem and absence of leptonic $CP$ phases are both linked to all Higgs VEVs being real. 

Though $CP$ is  broken softly by dimension 3  terms in~(\ref{eq:M_i}), the only  possible  $CP$ breaking term in the scalar potential with Higgs content of Table~\ref{Tab:Table1} is  $\mu^2 Tr \tilde{\phi}^\dagger  \phi + Hc.$   However $\mu^2$ is real due to $P$ and therefore does not break $CP$.   Since all terms in the Higgs potential are real, the  VEVs can be naturally real thus solving the strong $CP$ problem and predicting  the absence of leptonic $CP$ violation~\cite{Kuchimanchi:2010xs,Kuchimanchi:2012xb}...that is  the absence of tree-level Dirac neutrino phase and Majorana phases in the Pontecorvo-Maki-Nakagawa-Sakata (PMNS) matrix. The electron electric dipole moment (EDM) will also be unobservably small.


Another distinguishing feature of our model is that it can generate an observable neutron EDM (or equivalently $\bar{\theta}$) in large regions of parameter space. In section~\ref{sec:masses} we saw that the charge neutral $X$-particle would be automatically stable if  $M_X >> f v_R \sim f' v_R$.  
 In a similar region of parameter space in the quark sector, that is for $M > v_R$, as shown in~\cite{Kuchimanchi:2010xs,Kuchimanchi:2012xb}, the strong $CP$ phase $\bar{\theta}$ generated radiatively due to the up sector quarks is 
\begin{equation} 
\bar{\theta} \sim \sum_{\stackrel{i=1~to~3}{j=1~to~4}} \left({\frac{1}{16 \pi^2}}\right)   \ Im \left(\frac{{h^{d}_{i4}}{h^{u}_{4j}}h^{u}_{ji}}{h^d_{ii}}\right) \left[{\frac{v_R}{M}}\right]^2
\label{eq:theta_higgs}
\end{equation} 
where $h^{u,d}_{ij}$ are the up and down Hermitian Yukawa matrices which are $4 \times 4$ (owing to the presence of the 4th vectorlike family).  The above equation holds in the physical basis where the light $3 \times 3$ down sector is diagonal (see~\cite{Kuchimanchi:2010xs}). 

An important thing to note in the above equation is the suppression factor $[v_R/M]^2$.  Due to this, even if the Yukawa couplings involving the $4^{th}$ family are large like the third generation, $\bar{\theta}$ can be within its experimental bounds.  For example if the $v_R \sim 10^{14~to~15} GeV$ (a scale hinted at, though not required by the observed neutrino mass-squared splittings), and $M\sim M_{Pl} \sim 10^{19} GeV,$ then using equation~(\ref{eq:theta_higgs}) we can see that radiative corrections generate $\bar{\theta}\sim 10^{-10}$ to $10^{-12}$ if  $h^u_{44} \sim 1, h^u_{34} \sim e^{i\phi}, h^d_{34} \sim h^{d}_{33}$. This is not only within the present bounds but also could be detected in the ongoing and future neutron EDM experiments and is therefore exciting. This relevant example evaluating $\bar{\theta}$ in the region $M > v_R$, with well motivated mass-scales  and large fourth generation Yukawas was not provided in~~\cite{Kuchimanchi:2010xs,Kuchimanchi:2012xb}.


For the other case $M << v_R$, as shown in~\cite{Kuchimanchi:2010xs,Kuchimanchi:2012xb}, the suppression factor $[v_R/M]^2$ in Eq.~(\ref{eq:theta_higgs}) is replaced by the logarithmic factor $ln(v_R/M)$.  In this case so that $ \bar{\theta} < 10^{-10}$, the Yukawa terms involving the fourth generation must be smaller than or at best equal to the first generation Yukawas.  Moreover, an observable neutron EDM with $\bar{\theta} \geq 10^{-13}$ is expected to be generated~\cite{Kuchimanchi:2012xb}. This is interesting because the $X$-particle Yukawas with the scalar bi-doublet had to be similarly small so as to kinematically stabilize the charge-neutral fermion in a similar region  $M_X << f v_R \sim f'v_R$.

 Before we proceed to non-minimal models, we note that while there is no leptonic CP violation in the minimal strong CP solving model, $M_X$ in Eq.~(\ref{eq:dark}) can be complex and the dark sector can contain a soft $CP$ violating phase. 

 
\subsection{Non-minimal models} 

If we introduce an additional  vectorlike lepton doublet family ($L_{4L,4R}, L'_{4L,4R}$) and provide it a real intrinsic parity $\pm 1$, then it will  couple to the usual families through an equation analogous to Eq.~(\ref{eq:M_i}) with $Q \rightarrow L$ and $M_i \rightarrow M^L_i$. It is easy to see that $CP$ phases will be generated in the leptonic sector in this case,  in a manner similar to that of the quark sector.  This is essentially an extension in the fermionic sector that goes beyond what is minimally needed for dark matter and it gives rise to leptonic $CP$ violation. 

However even in this case the electron EDM does not get generated  if there is no new physics (other than possible dark matter physics) at low enough energy scale.  To see this, note that the neutron EDM gets generated in the quark sector through $\bar{\theta}$ given by Eq.~(\ref{eq:theta_higgs}) (with or without the suppression in the square brackets). Whether the factor in the square brackets is there or not, this equation is independent of the weak scale ($v_{wk}$), and in particular $\bar{\theta}$ is not suppressed by factors such as $v_{wk}/M$ or $v_{wk}/v_R$.    

However there is no analogous $\bar{\theta}$ term in the leptonic sector to generate an electron EDM.  Thus when new physics is high enough that it decouples, the electron EDM would not receive any additional contribution beyond the small unobservable amount expected in the standard model.

Thus with an additional vectorlike leptonic family with intrinsic parity $\pm 1$, $CP$ phases are expected to be present in the PMNS matrix but the electron EDM may still not be generated.  The presence of a measurable neutron EDM  and an absence of electron EDM is not only consistent with our model but is also expected in a large region of parameter space, and considering the fact that no new physics beond the standard model  has so far been discovered by the Large Hadron Collider (LHC).

On the other hand  adding more Higgs fields beyond those in the minimal model, does not introduce leptonic $CP$ violation even in the PMNS matrix. For example,  a singlet scalar~\cite{Kuchimanchi:2010xs}  or a second Higgs bi-doublet~\cite{Kuchimanchi:2012xb} can be added to break $CP$ spontaneously instead of softly. 
However their VEVs do not generate  the strong  $CP$ phase, and due to  the same reason do not also generate leptonic CP violation. Adding $X$-particles as dark matter as we have done in this work,  does not change this result as they do not couple to  leptons.  

 
\subsection{Prevailing view on leptonic CP violation} If $CP$ violation is hard in nature as it is in the standard model (through dimension 4  Yukawa couplings), then all $CP$ violating phases including all the leptonic phases are expected to be present at the tree level, since they would be generated during renormalization.  However the prevailing view in the field seems to be  more biased than this --  that whether $CP$ violation is hard \emph{or not}, it is expected that leptonic $CP$ phases are present.  

For example a recent review by Branco et al~\cite{RevModPhys.84.515} states:   \begin{quote} ``From a theoretical point of view, the complex phase in the CKM matrix
may arise from complex Yukawa couplings and/or from
a relative $CP$-violating phase in the vacuum expectation
values (VEV) of Higgs fields. \emph{In either case,} one expects
an entirely analogous mechanism to arise in the lepton
sector, leading to leptonic $CP$ violation (LCPV)." \end{quote} We have italicized some words for emphasis.   


Our prediction of absence of tree-level leptonic $CP$ violation in the minimal and some non-minimal models with $CP$ \emph{and} $P$ that we discussed thus differs from the  prevalent view. 
For completeness we now consider more general ways of making this prediction using $CP$ \emph{and} an additional symmetry.

\subsection{Other symmetries} In lieu of $P$ we now use other symmetries with $CP$ and get vanishing leptonic $CP$ violation by making slight changes to existing models: 
\begin{itemize}
	\item   A way to solve the strong $CP$ problem  using the Nelson-Barr mechanism~\cite{Nelson:1983zb,PhysRevLett.53.329}, and have neutrino mixing,  is the model by Branco et al.~\cite{Branco:2003rt} where a vector-like iso-singlet  down quark ($D_{L,R}$),  3 right handed neutrinos and a Higgs singlet $S$  have been added to the usual standard model.  $CP$ and $Z_4$ are imposed in~\cite{Branco:2003rt} so that under $Z_4$,  $D_{L,R}, S \rightarrow -D_{L,R}, -S$ and $L_{iL}, e_{iR}, \nu_{iR} \rightarrow iL_{iL}, ie_{iR},  i\nu_{iR}$, where the subscript $i = 1, 2, 3$ are the usual 3 generations.  All other fields such as the usual left-handed quark doublets and right-handed singlets are $Z_4$ invariant.   
	
	When $S$ picks a complex $CP$ violating VEV then owing to its Yukawa coupling  $D_{L} (h_i S + h'_i S^\star) d_{iR}$  with the usual 3 right-handed iso-singlet quarks $d_{iR}$, complex phases enter the quark mass matrix and the $CKM$ phase is generated.  However $S$ and $S^\star$ also have $Z_4$ invariant Majorana type Yukawa couplings with the right handed neutrinos  such as $\nu_{iR} C(f_{il}S +f_{il}'S^\star) \nu_{lR}$ that induce leptonic $CP$ violation.  Thus Branco et al find a common origin for the leptonic and quark sector $CP$ phases. 

However we note that if we change the transformation properties to make all leptons  invariant under $Z_4$, then $S$ would not couple to the leptons and there is no $CP$ violation generated in the leptonic sector, while the quark sector remains unaffected.  The neutrino mixing happens as terms such as $m_{il} \nu_{iR}C \nu_{lR}$ are now permitted. $m_{il}$ and all leptonic Yukawa couplings are real due to $CP$ and there is an absence of leptonic $CP$ violation.  Fermions ($X$) can be added and if $X \rightarrow i X$ under $Z_4$ they do not couple to leptons and can serve as stable dark matter.

In case of $P$  that we considered earlier the matter content of the minimal  model  unambiguously predicted the absence of leptonic $CP$ violation.  However for the minimal matter content with $Z_4$ (or $Z_2$ if dark matter is not introduced),  leptonic $CP$ violation is present or absent depending on how $Z_4$ (or $Z_2$) is imposed.   

\item
 We also consider the simple extension of the standard model (plus 3 right handed neutrinos) with two Higgs doublets $H_o$ and $H_e$, and where  we do not solve the strong $CP$ problem. 
 
 $Z_2$ is introduced so that under it,  $H_o \rightarrow -H_o$ and $H_e \rightarrow H_e$.  Thus $H_o$ and $H_e$ are $Z_2$ odd and even respectively.  Moreover one generation of quarks, say the first generation $q_{1L}, u_R, d_R$ are $Z_2$ odd while the remaining two quark generations and all the leptons are $Z_2$ even.  The usual $CP$ symmetry is imposed so that all parameters of the model are real, except that  we allow soft breaking of $CP$ and $Z_2$ by dimension two Higgs potential terms $\mu^2 H^\dagger_o H_e$, with $\mu^2$ complex.  
 
The VEV of neutral component of $H_e$ is taken to be real without loss of generality. The neutral component of $H_o$ picks up a complex CP violating VEV.

Both $H_o$ and $H_e$ have Yukawa couplings with the quarks that are $Z_2$ and $CP$ invariant. The Yukawa coupling parameters  are real due to $CP$. For example terms involving $H_o$ such as $h_{1c} \bar{q}_{1L} H_o c_R$, $h_{2u} \bar{q}_{2L} H_o u_R $ and $h_{1b} \bar{q}_{1L} \tilde{H}_o b_R$ that couple the first generation to the $c$ and $t$ quarks (as well as to $b$ and $s$)  are present. While the quark Yukawa couplings involving $H_e$, such as $h_{2c}\bar{q}_{2L} H_e c_R$, $h_{ii}\bar{q}_{iL} H_e u_{iR}$ (with $u_{iR} \equiv u, c, t$ for $i = 1, 2, 3$ respectively), $h_{ii}\bar{q}_{iL} \tilde{H}_e d_{iR}$   are also $Z_2$ invariant and present. 
 
 However since all the leptons are $Z_2$ even, only $H_e$ has Yukawa couplings with the Leptons. There are no Yukawa terms involving $H_o$ and the Leptons. Moreover due to presence of right-handed neutrinos, Dirac and Majorana terms are present for the neutrinos, and generate all the leptonic mixing angles.  These terms are all real due to $CP$.     
     
Since leptons do not couple to $H_o$,  this model will therefore once again lead to an absence of leptonic $CP$ violation.   But since both $H_o$ and $H_e$ have Yukawa couplings with the quarks and $H_o$ picks up a $CP$ violating VEV, the Jarlskog invariant is non-zero for the quarks,  as can be checked.  However while several models with 2 Higgs doublets have been considered (for a review please see~\cite{Branco:2011iw}), ones that lead to an absence of leptonic $CP$ violation while generating the needed CKM Matrix and allowing all the neutrino mixing angles  seem not to be studied so far.
\end{itemize}
	\section{Conclusion} We showed  that parity and quantum nature of the laws governing the  universe may be at the heart of dark matter stability. Dark matter can be matter with a relative purely imaginary intrinsic parity phase that cannot be removed through field or parity symmetry redefinitions.  If Majorana terms are not present either for leptons or dark sector,  there is  enough symmetry to  redefine $P$  and remove the relative purely imaginary intrinsic parity phase.    We thus predict that neutrinos and neutral dark matter fermions must have Majorana masses if dark matter is stable due to parity. Further $P$ with $CP$ solves the strong $CP$ Problem and predicts the vanishing of tree-level leptonic phases (in the minimal model).   If leptonic $CP$ phases in the PMNS matrix are not experimentally detected or are very small it would be consistent with $CP$ being spontaneously or softly broken \emph{and} there also being an additional hidden or softly broken symmetry in nature such as $P$, $Z_2$ or $Z_4$.  
	
Basically if the Lagrangian of nature has $CP$ symmetry then it must be violated softly and/or spontaneously to produce the CKM $CP$ violating phase.  If nature also has a second symmetry such as $P$, $Z_2$, or $Z_4$ (or a flavour symmetry) then this can ensure that while CKM phase is generated the strong $CP$ phase and leptonic $CP$ phases do not also get generated at the tree level. The same symmetry (for example $P$) that protects the strong $CP$ phase from getting generated can also protect the leptonic $CP$ phases from being generated. The same symmetry $P$ can also stabilize dark matter.  	

Moreover if leptonic CP violation is not detected, it would be a set back for axionic solutions, since the smallness of leptonic phases  would also have to be  explained, not just the strong CP phase.
	
Since leptonic $CP$ violation can be generated radiatively,  allowing at least a one-loop suppression, a conservative upper bound is that the induced leptonic phases in the PMNS matrix are less than $\delta_{ckm}/(16 \pi^2)  \sim 0.5^\circ$ from the $CP$ conserving values of  $0$ or $\pi$. In fact they will be much lesser as there will also be suppression of radiative corrections due to the increasing scale of new physics.  Currently experiments are being planned or underway~\cite{Huber:2009cw,Coloma:2012wq,Bayes:2012at} to achieve a sensitivity of about $5^\circ$. 

We also find that leptonic $CP$ phases can be present in the PMNS matrix in closely related or some non-minimal versions of these models, as $P, Z_2$ or $Z_4$ can then be imposed so that only the strong CP phase is protected from being generated at the tree level. However even in non-minimal cases,  in the axionless strong CP solving model that uses $P$ and $CP$ symmetries, the electric dipole moment of the electron will be immeasurably small if the scale of new physics is sufficiently high, while the neutron EDM can be generated at detectable levels due to radiative corrections to the strong $CP$ phase $\bar{\theta}$.

\appendix
\section{Inequivalent Lagrangians}
\label{appendix}

For the case discussed in subsection~\ref{sec:majorana} where the singlet Majorana fermion $X_M$ is given an intrinsic parity $\eta_{X_M} = i$,  we can provide either $\eta_L = \pm i$ or $\eta_L = \pm 1$ to the Leptons, so that under $P, L_{iL,iR} \rightarrow \eta_L L_{iR,iL}$.  Depending on the sign of $\eta^2_L$ this will lead to two inequivalent Lagrangians that conserve parity. The two Lagrangians will differ in the Majorana terms of the Leptons which for $\eta^2_L = 1$ will be as in Eq.~(\ref{eq:doublet_maj}) and for $\eta^2_L = -1$ will be given by
\begin{equation} 
\left({\frac{f_{ij}}{M}} \right)(L_{iL} H_L  L_{jL}H_L - L_{iR} H_R  L_{jR}H_R) + H.c. 
\label{eq:doublet_maj_i}
\end{equation}
Note that all the scalar fields have been assigned intrinsic parity 1.  There may be a doubt if the difference in the  sign occurring in the brackets of Eq.~(\ref{eq:doublet_maj}) and Eq.~(\ref{eq:doublet_maj_i})   will matter when physical quantities are calculated.   We now show that the two Lagrangians are physically inequivalent by examining the mass spectrum of the neutrinos for both cases, after the neutral components of the fields $H_{L}$ and $H_R$  pick up VEVS $v_L$ and $v_R$,  and $\phi$ picks up VEVs diag$\{\kappa, \kappa'\}$ (so that the weak scale $v_{wk} = \sqrt{|\kappa|^2 + |\kappa'|^2}$).  

To simplify the calculation we consider a toy model of $CP$ conserving Lagrangians with all real parameters, that can naturally have VEVs that are also all real.    Using $SU(2)_L$ and $SU(2)_R$ invariance without loss of generality, we can now set $v_L$ and $v_R$ to be positive. Making a further simplifying assumption that the Lepton generations do not mix,  the mass matrices of the $i^{th}$ generation of neutrinos now has the well known seesaw form 
$\left(\begin{tabular}{cc} $f_{ii} v'_L$ & $m_D$ \\ $m_D$ & $f_{ii} v'_R$  \end{tabular}\right)$ for $\eta^2_L = 1$ (due to Eq.~(\ref{eq:doublet_maj})), where $f_{ii}, m_D$ are real and $v'_{L,R} = v^2_{L,R}/M$ are real and positive. The Dirac mass term $m_D$ are real and can be obtained in terms of the real Yukawas $h_{ii}, \tilde{h}_{ii}$ of Eq.~(\ref{eq:dirac}) and real VEVs $\kappa, \kappa'$.    

For $\eta^2_L = -1$ the seesaw matrix is derived from eqs.~(\ref{eq:dirac}) and~(\ref{eq:doublet_maj_i}) and is slightly different, as it is $\left(\begin{tabular}{cc} $f_{ii} v'_L$ & $m_D$ \\ $m_D$ & $- f_{ii} v'_R$ \end{tabular}\right).$      
   
The two seesaw matrices will have different eigenvalue spectra and therefore they describe different physics. With a little work we can see that for the $i^{th}$ generation, the difference in the squares of mass eigenvalues of the heavy and light neutrinos is not exactly the same for the two cases. Flipping the sign of $f_{ii}$ or $m_D$ in either of the matrices  does not change the situation.

For the case where $\eta_{X_M} = i$ and $\eta_L = \pm 1$, parity stabilizes dark matter.  On the other hand if  $\eta_{X_M} = \eta_L = i,$ a $Z_2$ symmetry would have to be imposed to keep $X_M$ stable.   

A similar result can be obtained for the model in subsection~\ref{sec:xparticles} by calculating the mass spectrum of both the neutral $X-$particles and neutrinos for Lagrangians obtained using $\eta^2_X = \eta^2_L$ and $\eta^2_X = -\eta^2_L$.


\section{Dirac terms and stability due to P}
\label{sec:appendixb}

If symmetry under $X_R \rightarrow -X_R$ and $X'_L \rightarrow -X'_L$ is exactly  imposed so that Dirac type Yukawa couplings $h = h' = \tilde{h} = \tilde{h}' = 0$ in Eq.~(\ref{eq:dark}), it is easy to see that that the purely imaginary intrinsic parity phase can be defined away and $P$ does not stabilize dark matter.   This is because a field redefinition $X_R \rightarrow i X_R$ and $X'_L \rightarrow i X'_L$ can now change the relative minus to a plus sign in the brackets of the terms corresponding to $f$ as well as $f'$ (with $f' \rightarrow -f'$) in Eq.~(\ref{eq:dark}), making them similar to the plus sign in the Majorana term of the Leptons in Eq.~(\ref{eq:maj}).  All other terms remain unchanged.  The plus sign in these brackets implies that the intrinsic parities of both the Leptons and $X-$particles are $\eta_L = \eta_X = 1$ and there is no relative imaginary intrinsic parity phase. Thus  some Dirac type  Yukawa terms must be present in Eq.~(\ref{eq:dark}) if $P$  stabilizes dark matter, causing the charged fermion masses to necessarily split in section~\ref{sec:masses}. 

Note that if Dirac Yukawa couplings such as $h, \tilde{h} \neq 0$, the above field redefinition will also require $h, \tilde{h} \rightarrow ih, i\tilde{h}$  in Eq.~(\ref{eq:dark}).  However as mentioned in section~\ref{sec:xparticles}, $h, \tilde{h}$ are real   due to $P$ symmetry and the imaginary phase $i$ cannot be absorbed in $h$ and $\tilde{h}$. 


\providecommand{\href}[2]{#2}\begingroup\raggedright\endgroup

\end{document}